\input epsf
\input lecproc.cmm
\def\rcomp{\rho_{\rm c}}
\def\tacc{t_{\rm acc}}
\def\tesc{t_{\rm esc}}
\def\diff{{\rm d}}
\def\numax{\nu_{\rm max}}
\def\gammamax{\gamma_{\rm max}}

\contribution{Particle acceleration 
at shocks in relativistic jets}
\author{
J.G. Kirk}
\address{Max-Planck-Institut f\"ur Kernphysik,
Postfach 10 39 80, D-69029 Heidelberg, Germany}

\abstract{The theory of 
particle acceleration at shock fronts is briefly reviewed, 
with special emphasis on the production of the particles 
responsible for the nonthermal emission from blazars. The flat radio/IR 
spectra of these sources cannot be produced by diffusive acceleration 
at a simple nonrelativistic shock front propagating in a homogeneous medium. 
It can, however, 
be produced by a single unmodified mildly relativistic 
shock, if the pressure in the shocked gas is provided by the 
leptonic component, or, independently 
of the equation of state, by a relativistic shock which is  
oblique to the magnetic field. The analytic 
theory of these shocks makes several simplifications, but Monte-Carlo 
simulations exist which extend the range of validity. Of particular 
interest is acceleration in a tangled magnetic field. 
Here, however, the Monte-Carlo simulations have not yet yielded unambiguous
results. 
The \lq homogeneous\rq\ models of blazar 
emission are discussed, and is it shown that they imply a geometry of 
the emitting region which is laminar in form, with an aspect ratio of 
$d/R\le 3\%$ in the case of Mkn~421. 
Identifying these with relativistic shock fronts, a 
model of acceleration is described, which displays 
characteristic variations in the synchrotron spectral index with 
intensity.}

\titlea{1}{Introduction}
The problem of how particles are accelerated to nonthermal energies in 
relativistic jets has been discussed for a number of years (e.g., Begelman,
Blandford \& Rees~1984). Diffusive acceleration at 
nonrelativistic shocks is a possibility which 
appears to provide a reasonable picture of 
acceleration at several jet hot-spots which emit optical synchrotron 
radiation (Meisenheimer et al.~1989). However, the rapid variability of emission
from blazar jets implies substantial doppler boosting, which
suggests that the nonrelativistic theory may be inappropriate.
Relativistic shocks have also been known for 
some time to be potentially effective accelerators (Kirk \& 
Schneider~1987), but, unlike their nonrelativistic 
counterparts, they do not lead to a unique spectral index which is  
independent of the details of the scattering process and the 
shock speed. Instead, a wide range of spectral slopes is possible. 

The best developed model of blazar spectra is based on the picture of a 
shock front moving down a jet (Marscher \& Gear 1985). 
Both the magnetic field and the particle distribution vary with position, and
the observed radiation is a superposition of emission from 
different parts of the 
jet (Ghisellini, Maraschi \& Treves~1985,
Ballard et al.~1990, Hughes, 
Aller \& Aller~1991, Maraschi, Ghisellini \& Celotti~1992,
Levinson \& Blandford~1995, Marscher \& Travis~1996).
These models are generally called \lq inhomogeneous\rq. However,
recent results concerning the rapid variability of blazar sources at all 
frequencies 
(Wagner \& Witzel~1995) 
and the detection of their emission at energies of up to 
at least 
$10\,{\rm TeV}$ (Aharonian et al.~1997) have provided new 
restrictions on the possible acceleration mechanisms. 
Especially the observed simultaneous variations in X-rays and TeV gamma-rays
indicate that a single population of particles in a relatively localised region
 is responsible. 
Consequently, \lq homogeneous\rq\ models have been widely discussed
(Dermer \& Schlickeiser~1993, Macomb et al.~1995, 1996, 
Bloom 
\& Marscher~1996, Ghisellini, 
Maraschi \& Dondi~1996, Inoue \& Takahara~1996, 
Stecker, de~Jager \& Salamon~1996,
Comastri et al.~1997, 
Mastichiadis \& Kirk~1997) 
and there is rough agreement on the parameters of the
emission region.

Simultaneous variability of emission at widely differing 
wavelengths is more easily interpreted in terms of 
directly accelerated electrons than 
in the hadronic models
(Mannheim~1993, Mannheim et al.~1996), in which
the gamma-ray emission 
stems originally from ultra-relativistic protons, 
and only the optical and (in some sources) X-ray photons arise 
from directly accelerated electrons. 
Protons and electrons are accelerated in 
regions of different spatial extent 
on very different timescales, because the waves which provide the 
scattering centres have a wavelength comparable to the particle's gyro radius,
and the acceleration time of a particle is probably somewhat longer than its 
gyro period.
Consequently, one would expect the radiation produced by relatively 
low energy (GeV to TeV) electrons to have different 
variability properties to that produced by protons of energy above 
$1\,{\rm EeV}$. 
However, further information concerning the emission of 
these objects at energies above $20\,{\rm TeV}$ 
(Meyer \& Westerhoff~1996) may well provide the most reliable way 
distinguishing between the leptonic and hadronic models.


\titlea{2}{Basic properties of shock acceleration}
It is well-known that particles which diffuse in the neighbourhood of a 
nonrelativistic shock front can be accelerated into a power-law 
distribution such that the density of particles with Lorentz factor 
$\gamma$ is a power-law, $n(\gamma)\propto \gamma^{-s+2}$, with 
$$
s=3\rcomp/(\rcomp-1)
\eqno{(1)}
$$
where $\rcomp$ is the compression ratio of the 
shock front. 
The physical basis of acceleration is that an energetic 
particle scatters elastically off magnetic fluctuations in
the background plasma, which enables it to cross and recross the shock front.
Simple kinematics lead to an energy gain on each crossing, since, at a
shock front, the scattering centres take part in the plasma compression. This
competes with the possibility that a particle moves off into the downstream 
plasma and does not return, to yield a power-law spectrum (for a review
see Kirk, Melrose \& Priest~1994). 

Before they have had 
time to cool, the synchrotron emission of electrons accelerated into a
distribution described by Eq.~(1) is a power law $I(\nu)\propto \nu^{-\alpha}$, with $\alpha=(s-
3)/2$. Thus, a strong nonrelativistic 
shock front in a fully ionised gas, which has 
$\rcomp=4$, produces $\alpha=0.5$. There are several effects which 
modify this result. However, most of them lead to 
a softening of the predicted spectrum, at least for lower energy 
electrons (e.g, Bell~1987, 
Ellison \& Reynolds~1991, Duffy, Ball \& Kirk~1995).
In the case of blazars, the homogeneous model requires a harder spectrum 
-- for Mrk~421, for example, $\alpha\approx0.35$. 

\begfig 0 cm
\centerline{
\hbox{
\epsfxsize=6 cm
\epsffile{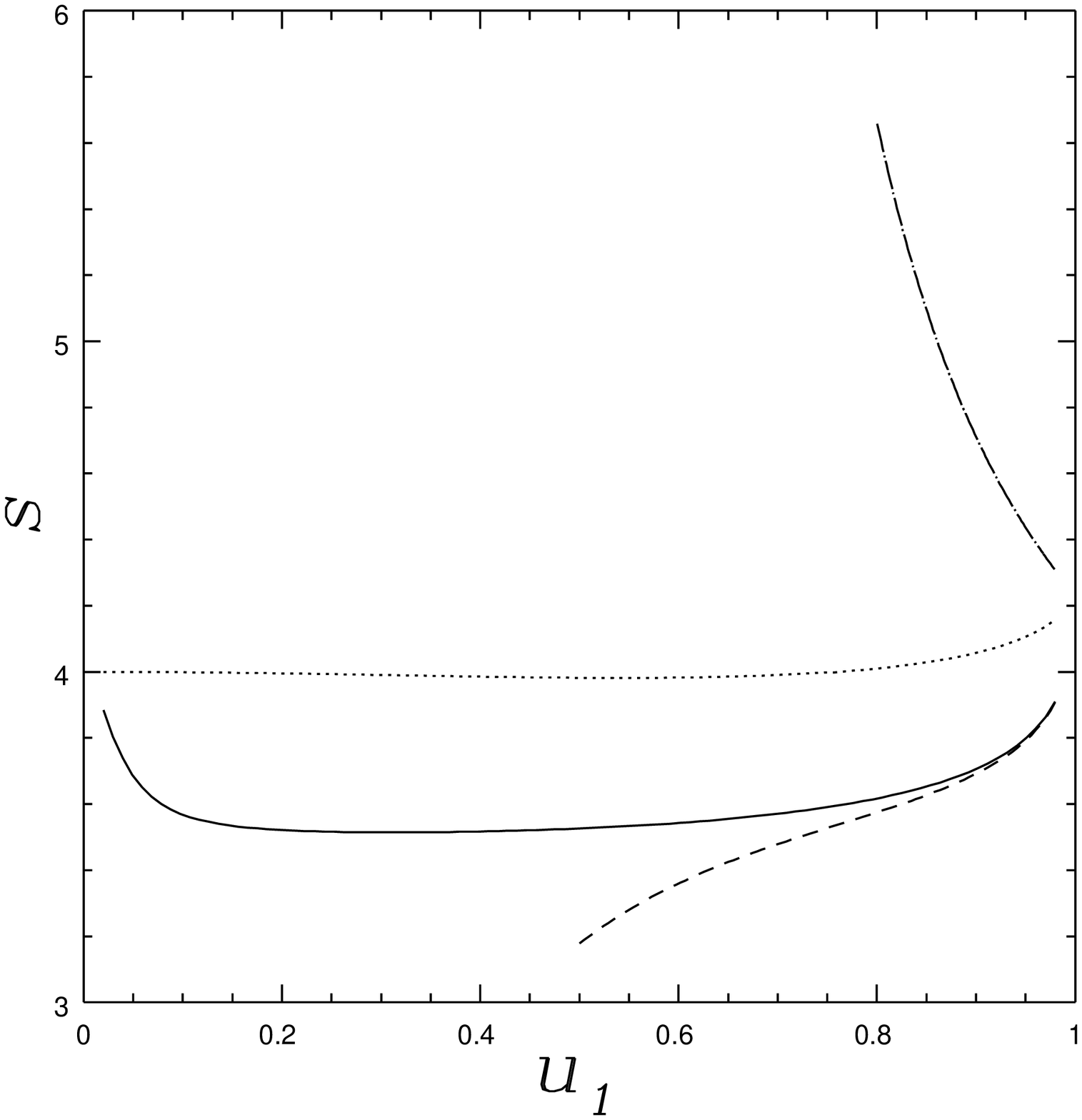} 
\epsfxsize=6 cm
\epsffile{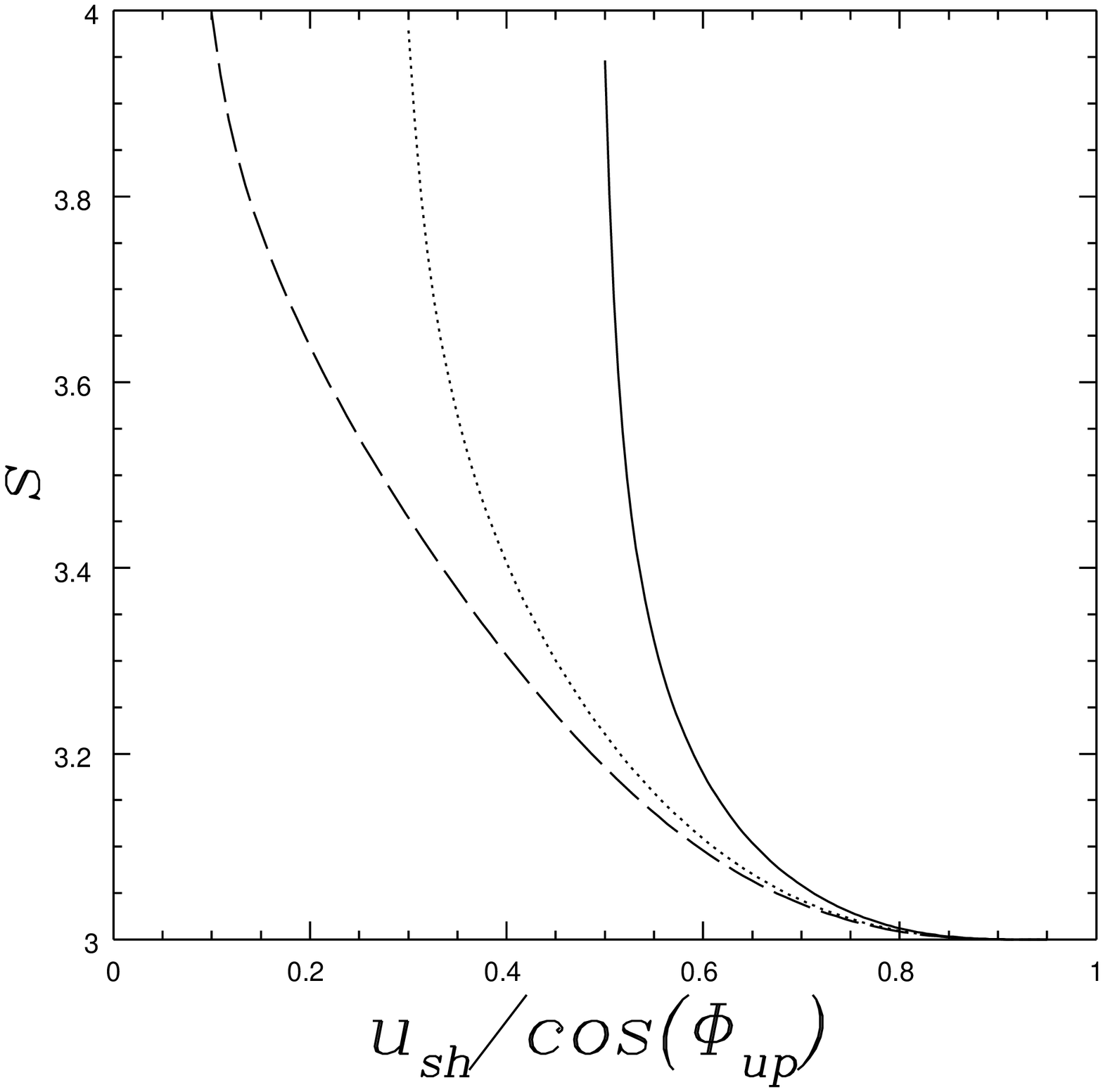} 
}}
\figure{1}{The power law index of particles accelerated at a relativistic shock
front. The left-hand panel shows the index $s$ of the phase space-density for a
parallel shock as a function of the velocity of the shock seen from the rest
frame of the upstream plasma. Four different equations of state are used: the
dashed-dotted line is for a relativistic gas, the dotted line for a gas in
which the ions provide the pressure (i.e., hot ions, cold electrons), the solid
line is for hot electrons and cold ions, and the dashed line is for a gas
containing a population of electron positron pairs. A relatively hard spectrum
is obtained for intermediate speeds if the electrons are hot, or 
pairs present. At high speeds ($\Gamma>5$) all curves tend towards the value
$s=4.2$. In the right-hand panel, the effect of a finite angle $\Phi_{\rm up}$
between the
shock normal and the magnetic field is shown, for three different values of the
upstream shock speed $u_{\rm sh}=0.1$, $0.3$ and $0.5$ and a compression ratio
of 4. As the speed of the intersection point of the shock and field line
($=u_{\rm sh}/\cos(\Phi_{\rm up})$) 
approaches $c$, the spectra harden.}
\endfig

Blazar jets are almost certainly in relativistic motion, 
but although relativistic shocks offer a range of indices, the required 
$s=3.7$ is not easily obtained. As shown in Fig.~1, mildly 
relativistic shocks ($0.1<u_1<0.9$, with $u_1$ the speed of the shock seen from
the rest frame of the upstream plasma) give this value
provided the electron pressure is important behind the shock front. This
effect is due not to a radical change in the way in which acceleration
operates, but simply to the increased compression ratio across the shock front when
electrons are heated to a temperature comparable with their rest-mass. 
Whether the processes mediating a relativistic shock lead to such 
heating is unknown. However, it seems unlikely that the background 
plasma, which is assumed to carry more energy flux than the accelerated
component, should remain invisible, despite having an electron temperature
of several MeV.
A more attractive alternative is presented by 
oblique relativistic
shocks. In contrast to the 
nonrelativistic case, where the obliquity has no effect on the spectral 
index (Axford~1980), oblique relativistic shocks show much 
harder spectra (Kirk \& Heavens~1989), ranging up to 
$s=3$. The reason for this behaviour lies in the increased importance of 
reflections from the shock front as a fundamental accelerating process. If 
particles are tightly bound to field lines, and are also randomly 
distributed in gyro-phase -- two assumptions behind the analytic 
treatment of Kirk \& Heavens~(1989) -- the probability 
that they are reflected by the magnetic compression at the shock front 
is significant. As the shock speed increases, the energy gained on 
reflection goes up, since the effective speed of the \lq mirror\rq\ is 
that of the intersection point of a magnetic field line with the shock 
surface. The overall effect is to harden the spectra as shown in 
Fig.~1. 

It has been pointed out that this type acceleration might be 
supressed in highly relativistic flows (Begelman \& Kirk~1990).
In the absence of cross-field transport, crossing and recrossing requires that
a particle move faster along a field line than the intersection point of that
line
with the shock front. This speed will normally exceed $c$ for a highly
relativistic shock, unless the upstream field is aligned to within an angle
of $1/\Gamma$ of the shock normal. (Here $\Gamma=(1-u_1^2/c^2)^{-1/2}$ 
is the Lorentz factor
associated with the 
speed of the shock seen from the rest frame of the upstream plasma.)
As a result, recrossing of the shock into the upstream medium would not 
be permitted and acceleration would be limited to a single 
encounter -- a process usually referred to as \lq shock-drift\rq\ 
acceleration. 

The two main assumptions used in the analytic theory of relativistic 
oblique shocks (conservation of magnetic moment, and 
absence of cross-field transport) are readily relaxed in a 
Monte-Carlo simulation, and 
this program has been carried out by several groups (Ostrowski~1991,
Lieu et al.~1994, Naito \& Takahara~1995). The simulations show
that the degree of cross-field transport is critical. The analytic 
results are reproduced if it is small, but the spectra soften to reach 
the value for a parallel shock once the transport properties become 
isotropic. Another feature emphasised by these simulations is that
the spectrum produced depends upon the type of scattering used in the 
simulation -- a property also of parallel relativistic 
shocks (Kirk \& Schneider~1988).   

Scattering is ultimately just one way of describing 
the perturbed trajectory of a particle 
which encounters fluctuations of the magnetic field about its average 
value. If, as would appear from the above work, the obliquity of the 
shock front is crucial, fluctuations might have an important role in 
changing the angle with which a particular field line hits the shock 
front, a process which is not included in simulations which simply 
superpose stochastic scattering on motion in a homogeneous field.
To investigate this 
effect, two groups have performed simulations in which a 
tangled magnetic field with a
random component is realised explicitly (Ballard \& 
Heavens~1992, Ostrowski~1993). 
Assuming that the field is frozen into the plasma, which supports a 
shock front consisting of a simple velocity discontinuity, the particle 
orbits are followed by numerically integrating the equations of 
motion. From the considerations mentioned above, we might expect that 
reflection from the shock front plays the dominant role in such a 
situation. In fact, a tangled field will always inhibit the escape of a 
particle reflected off the advancing edge of the shock front and cause 
it to be caught once again. Thus, a layer of energised particles may 
well be built up on the upstream side of the shock front. On 
transmission to the downstream plasma, tangling of the field will in 
principle permit some particles to return to the shock front. In a 
highly oblique shock, this process is unimportant, which may
also be the case at a relativistic shock with tangled fields.
However, this is mere speculation; the two Monte-Carlo simulations
unfortunately yield conflicting results, for 
reasons which have not yet been resolved. A possible explanation is 
that the techniques by which the stochastic field is realised are 
different in each simulation, and it is not clear 
that the statistical properties of the 
resulting field lines are equivalent.
 
Stochastic magnetic fields have recently been investigated in connection with  
cosmic ray transport (Chuvilgin \& Ptuskin~1993) and 
with nonrelativistic shocks (Duffy et al.~1995,
Kirk, Duffy \& Gallant~1996), where they have been shown 
to be one of the few 
effects capable of modifying the rather robust result given by 
Eq.~(1). Simulation techniques which exploit this 
approach (Gieseler et al.~1997) may prove useful in 
resolving the question for relativistic shocks. 

\titlea{3}{A model of the time dependence of acceleration}
In the case of the homogeneous synchro-self-compton model, the following simple
estimate can be made (cf.~Takahashi et al.~1996). Denote by $\nu_{10}$ the
highest energy photons emitted as synchrotron radiation in units of $10\,$keV, 
and by $t_3$ the
variability timescale of this emission in thousands of seconds, which we
identify 
with the cooling time 
by synchrotron radiation. Then
$\nu_{10}\approx 10^{-12}\gamma^2 B \delta$ and
$t_3\approx 10^6\gamma^{-1} B^{-2}\delta^{-1}$, where $\gamma$ is the Lorentz
factor of the particle in the rest frame of the emitting plasma, $B$ the
magnetic field in gauss and $\delta$ the Doppler boosting factor. These
relations imply
$B=1\times t_3^{-2/3}\nu_{10}^{-1/3}\delta^{-1/3}\,$gauss and 
$\gamma=10^6 \times t_3^{1/3}\nu_{10}^{2/3}\delta^{-1/3}$.
Those objects which show TeV emission generally have roughly the same
luminosity in the synchrotron and the inverse compton parts of the 
spectrum,
which means that the energy densities of photons and magnetic field in the source
are of comparable magnitude. Setting them equal leads to an expression
for the \lq aspect ratio\rq\ of the source region $\eta=d/R$, where
$d=1000t_3\delta/c$ is a characteristic thickness measured
in the rest frame of the source, and $R$ is defined such that 
$\pi R^2$ is the area of the source when 
projected onto the plane of the sky. For Mkn~421, which has an apparent
luminosity of $6\times10^{44}\,{\rm erg\,s^{-1}}$, one finds
$\eta=7\times 10^{-5} \nu_{10}^{-1/3} t_3^{1/3}\delta^{8/3}$, and, 
inserting $\delta=10$, $\nu_{10}=1$ and $t_3=1$, we find 
$\eta\approx 0.03$. 
This indicates that we are dealing with an essentially two-dimensional source
region, which it is tempting to identify with a relativistic shock front. 

The observation of electrons during their acceleration by the shock 
front of SN1987A (Staveley-Smith et al.~1992) 
led to the development of a simple 
but useful model for the time-dependence of shock acceleration (Ball \& 
Kirk~1992) which can be applied to the present case 
simply by generalising the kinematics to relativistic 
flows, and 
including the effects of synchrotron energy losses.
The basic idea is to divide the system up into two regions: one 
-- which we can imagine to be around or just in front of the shock front 
-- in which 
particles are either repeatedly reflected, or cross and recross the front 
and so undergo continuous acceleration at a rate $\tacc^{-1}$, and a second
which is the downstream plasma, 
where there is no further
acceleration, but only cooling. Escape 
from the acceleration zone into the downstream plasma 
occurs at a rate $\tesc^{-1}$. 
The number $N(\gamma)\diff \gamma$ of 
particles in the acceleration zone with Lorentz factor between $\gamma$ and 
$\gamma+\diff\gamma$ is governed by the equation
$$
 {\partial N\over \partial t} + {\partial\over\partial \gamma} 
\left[\left(
   {\gamma\over\tacc}  -\beta_{\rm s}\,\gamma^2 \right) N \right] +
   {N\over \tesc}  = Q \delta (\gamma - \gamma_0)
\eqno{(2)}
$$
(e.g., Kirk, Melrose \& Priest~1994),
where    
$\beta_{\rm s}= 
4\sigma_{\rm T}/(3m_{\rm e} c)(B^2/8\pi)$
with $\sigma_{\rm T}= 6.65 \cdot 10^{-25} {\rm cm}^2$ the 
Thomson cross section. The first term in brackets in 
Eq.~(2) describes acceleration, 
the second describes the rate of energy loss due to synchrotron radiation 
averaged over pitch-angle in a magnetic field $B$ (in gauss). Particles 
are assumed to be picked up (injected) into the acceleration 
process with Lorentz factor $\gamma_0$ at a rate $Q$ particles per 
second.    

In a one-dimensional picture, with the $x$ coordinate along the shock normal 
the kinetic 
equation governing the differential 
density $\diff n(x,\gamma,t)$ of particles which have escaped into the 
downstream plasma and are in  
the range $\diff x$, $\diff \gamma$ is
$$
{\partial n\over \partial t} -
{\partial\over \partial \gamma}(\beta_{\rm s}\,\gamma^2\,n) =
{N(\gamma,t)\over\tesc}\delta(x-x_{\rm s}(t)) 
\eqno{(3)}
$$
where $x_{\rm s}(t)$ is the position of the shock front at time $t$,
and a coordinate system has been used which in which the plasma is at 
rest. Note that the \lq injection\rq\ term in this equation is provided 
by those particles escaping the acceleration zone.

Equations (2) and (3) are simple to solve. Using the 
synchrotron Green's function the electron 
distribution is then readily linked to the emitted radiation,
given the time dependence of the rate 
$Q$ at which particles are picked up by the acceleration process
(Kirk, Rieger \& Mastichiadis~1997). 
For example,
a simple representation of a flare is found by setting
$Q(t)=Q_0$ for $t<0$ and $t>10\tacc$ and $Q(t)=2 Q_0$
for $0<t<10\tacc$

\begfig 0 cm
\centerline{
\hbox{
\epsfxsize=6 cm
\epsffile{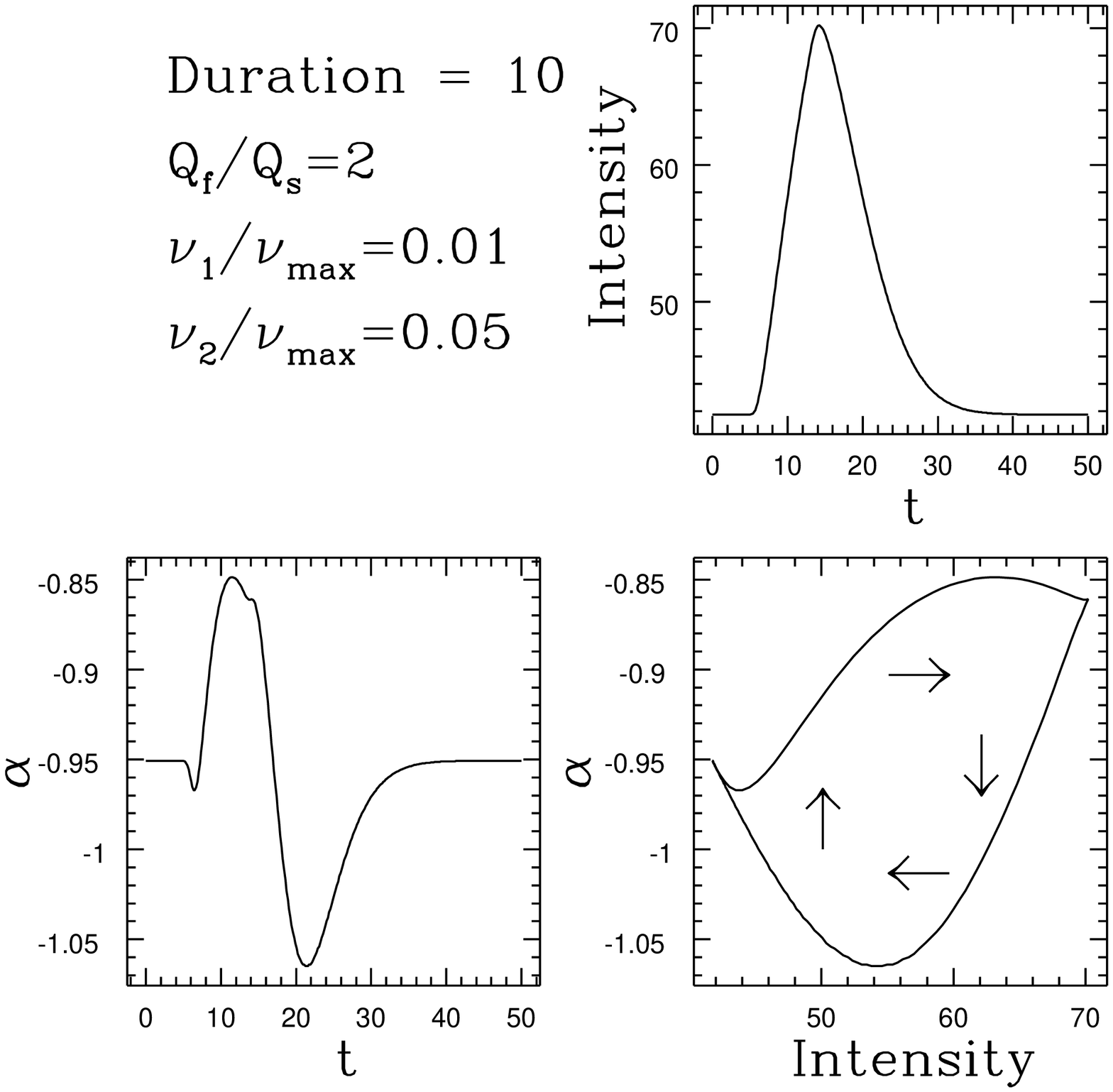} 
\epsfxsize=6 cm
\epsffile{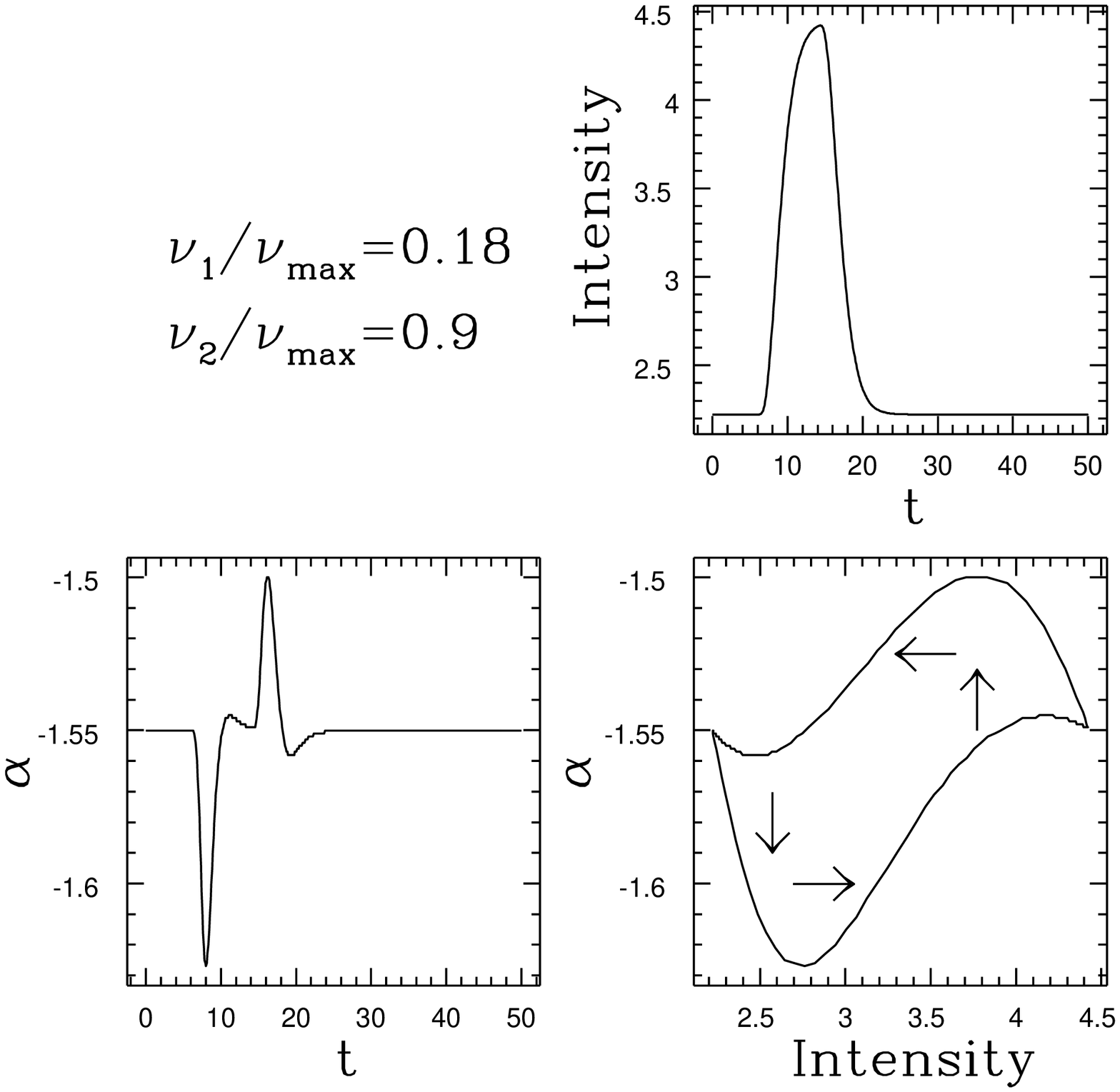} 
}}
\figure{2}{
The predicted 
behaviour of the synchrotron emission and spectral index during a flare.
In the left-hand set of three panels, the light curve, and spectral index
at low frequency ($\nu=\numax/20$) are shown (with the index $\alpha$
calculated from the flux ratio at $\nu$ and $\nu/10$). 
The loop in the $\alpha$ vs.\
intensity plane is traversed in the clockwise direction. In the right-hand
panels, 
the same plots are shown at $\nu=0.9\numax$. Here the loop is followed in the 
anti-clockwise direction.}
\endfig

The corresponding synchrotron emission in the rest frame of the source 
is shown in Fig.~2.
The behaviour of the spectrum at frequencies well below the maximum,
where the acceleration time $\tacc$ is much shorter than the synchrotron
cooling time $(\beta_{\rm s}\gamma)^{-1}$ shows the
characteristic \lq slow-lag\rq\ observed in several sources
(e.g., Pks~2155--3034: Sembay et al.~1993 and Mkn~421: 
Takahashi et al 1996). This can be interpreted simply in
terms of synchrotron cooling (Tashiro et al.~1995).
However, closer to the maximum frequency, the finite acceleration time comes
into play (At $\gamma=\gammamax$, 
we have $\tacc=(\beta_{\rm s}\gammamax)^{-1}$.)
Then it is possible to produce hysteresis curves which are followed in the
opposite direction, namely anti-clockwise. 
This behaviour is characteristic of the acceleration mechanism. Although the
details of such loops will depend on factors neglected here such as the energy
dependence of $\tacc$, inhomogeneities in the source geometry, and smoothing 
effects produced by the finite light travel time across the face of the source,
the conclusion remains 
that anti-clockwise motion is the signature of the acceleration
mechanism. Combined with the fact that rapid variation implies 
laminar source geometry and that substantial doppler boosting is required, 
this may also be the signature of 
a relativistic shock front.

\begrefchapter{References}
\ref
Aharonian F. et al.\ 1997 submitted to A\&A
\ref
Axford W.I. 1980 Proc.\ 17th.\ Int.\ Cosmic Ray Conf.\ (Paris) 12, 155
\ref
Ball L., Kirk J.G. 1992 ApJ 396, L39
\ref
Ballard K.R., Mead A.R.G., Brand P.W.J.L., Hough J.H. 1990 MNRAS 243, 640
\ref
Ballard K.R., Heavens A.F. 1992 MNRAS 259, 89
\ref
Begelman, M.C., Blandford R.D., Rees M.J. 1984 Reviews of Modern Physics 56,
255
\ref
Begelman M.C., Kirk J.G. 1990 ApJ 353, 66
\ref
Bell A.R. 1987 MNRAS 225, 615
\ref
Bloom S.D., Marscher A.P. 1996 ApJ 461, 657
\ref
Chuvilgin L.G., Ptuskin V.S. 1993 A\&A 279, 278
\ref
Comastri A., Fossati G., Ghisellini G., Molendi S. 1997 ApJ 480, 534
\ref
Dermer C.D., Schlickeiser R. 1993 ApJ416, 458
\ref
Duffy P., Ball L., Kirk J.G. 1995 ApJ 447, 364
\ref
Duffy P., Kirk J.G., Gallant Y.A., Dendy R.O. 1995 A\&A 302, L21
\ref
Ellison D.C., Reynolds S.P. 1991 ApJ 382, 242
\ref
Ghisellini G., Maraschi L., Dondi L. 1996 A\&A Supplement 120C, 503
\ref
Ghisellini, G., Maraschi, L., Treves, A. 1985, A\&A 146, 204
\ref
Gieseler U.D.J., Duffy P., Kirk J.G., Gallant Y.A. 1997 Proc.\
25th.\ Int.\ Cosmic Ray Conf.\ (Durban)
\ref
Hughes P.A., Aller H.D., Aller M.F. 1991 ApJ 374, 57
\ref
Inoue S., Takahara F. 1996 ApJ 463, 555
\ref
Kirk J.G., Duffy P., Gallant Y.A. 1996 ApJ 314, 1010
\ref
Kirk J.G., Heavens A.F. 1989 MNRAS 239, 995
\ref
Kirk J.G., Rieger F.M., Mastichiadis A. 1997 in preparation
\ref
Kirk J.G., Melrose D.B., Priest E.R. 1994 \lq Plasma Astrophysics\rq\
Springer-Verlag, New York
\ref
Kirk J.G., Schneider P. 1987 ApJ 315, 425
\ref
Kirk J.G., Schneider P. 1988 A\&A 201, 177
\ref
Lieu R., Quenby J.J., Drolias B., Naidu K. 1994 ApJ 421, 211
\ref
Levinson A., Blandford R.D. 1995 ApJ 449, 86
\ref
Macomb, D.J. et al. 1995 ApJ 449, L99
\ref
Macomb, D.J. et al. 1996 ApJ 459, L111 (Erratum)
\ref
Mannheim K. 1993 A\&A 269, 67
\ref
Mannheim K., Westerhoff S., Meyer H., Fink H.-H. 1996 A\&A 315, 77
\ref
Maraschi F., Ghisellini G., Celotti A. 1992 ApJ 379, L5
\ref
Marscher A.P., Gear W.K. 1985 ApJ 298, 114
\ref
Marscher A.P., Travis J.P. 1996 A\&A Supplement 120C, 537
\ref
Mastichiadis A., Kirk J.G. 1997 A\&A 320, 19
\ref
Meisenheimer K., R\"oser H.-J., Hiltner P.R., Yates M.G., Longair M.S., Chini
R., Perley R.A. 1989 A\&A 219, 63
\ref
Meyer H., Westerhoff S. 1996 in \lq Proceedings of the Heidelberg Workshop on 
Gamma-ray emitting AGN\rq, Eds. J.G.~Kirk, M.~Camenzind, C. von~Montigny,
S.~Wagner, MPI-Kernphysik internal report No. MPI H-V37-1996, page 39
\ref
Naito T., Takahara F. 1995 MNRAS 275, 1077
\ref
Ostrowski M. 1991 MNRAS 249, 551
\ref
Ostrowski M. 1993 MNRAS 264, 248
\ref
Sembay S., Warwick R.S., Urry C.M., Sokoloski J., George I.M.,
Makino F., Ohashi T., Tashiro M. 1993 ApJ 404, 112
\ref
Staveley-Smith L. et al. 1992 Nature 355, 147
\ref
Stecker F.W., de~Jager O.C., Salamon M.H. 1996 ApJ 473, L75
\ref
Takahashi , Tashiro, Madejski G., Kubo H., Kamae T., Kataoka J., Kii T., Makino
F., Makishima K., Yamasaki N. 1996 ApJ 470, L89
\ref
Tashiro et al.\ 1995 PASJ, 47, 131
\ref
Wagner S.J., Witzel A. 1995 ARA\&A 33, 163
\endref
\bye